\newcommand{\beq}{\begin{equation}}
\newcommand{\eeq}{\end{equation}}

\newcommand{\id}
 {i\kern.06em\hbox{\raise.25ex\hbox{$/$}\kern-.60em$\partial$}}

\newcommand{\bs}{/\kern-.52em b}
\newcommand{\qs}{/\kern-.52em s}

\newcommand{\p}{\partial}

\newcommand{\dd}
{\kern.06em\hbox{\raise.25ex\hbox{$/$}\kern-.60em$\partial$}}

\documentstyle[12pt]{article}
\begin{document}
\title{Exact Results of Strongly Correlated Systems at
Finite Temperature
\thanks{On leave of absence from the Physics Department,
Shanghai University, 201800, Shanghai, China}}
\author{{Sze-Shiang Feng}\\
1,{\small {\it High Energy Section, ICTP, Trieste, 34100, Italy}}
\\e-mail:fengss@ictp.trieste.it\\
2,{\small {\it CCAST(World Lab.), P.O. Box 8730, Beijing 100080}}\\
3,{\small {\it Modern Physics Department, University of Science and Technology
of China, 230026, Hefei, China}
}\\e-mail:zhdp@ustc.edu.cn}
\maketitle
\baselineskip 0.3in
\begin{center}
\begin{minipage}{135mm}
\vskip 0.3in
\baselineskip 0.3in
\begin{center}{\bf Abstract}\end{center}
  {Some rigorous conclusions of the  Hubbard model
  , Kondo lattice model and periodic Anderson
  model at finite temperature are acquired employing the
  fluctuation-dissipation
  theorem and particle-hole transform. The main conclusion
  states that for the three models,
  the expectation value of ${\bf \tilde{S}}^2-{\bf \tilde{S}}^2_z$
  will be of order $N_{\Lambda}$ at any
   finite  temperature  .
  \\PACS number(s): 75.10.Lp,71.20.Ad,74.65.+n,74.20.Mn
   \\Key words: strongly correlated, ODLRO,finite temperature}
\end{minipage}
\end{center}
\vskip 1in
Hubbard model(HM), Kondo lattice model
(KLM) and periodic Anderson model
(PAM)
are three typical  strongly-correlated electrons systems under
currently intensive investigations. They exhibit unusual
thermodynamic, magnetic and transport propertities (high-$T_c$)
\cite{s1}-\cite{s16}.
Dispite their superficial simplicities, exact
results about them are unusually difficult to obtain in more than
one dimensions \cite{s17}.
Fortunately, a series of rigorous results about the ground state of
the three models have been acquired
\cite{s1}-\cite{s16}.
To understand the magnetic properties of the ground state
of the models, the reflection symmetry, or up-down symmetry
is often imployed and this method was initially utilized
by Lieb\cite{s1}.
Having knowing some features in the ground states, one
is expecting some exact knowledge of the model at
{\it  finite}
temperature .
Apart from the various magnetic
properties of the models, an extremely interesting feature is that
it may provide an understanding of the high-$T_c$ superconducting
supported by cuprates such as YBaCuO. To investigate this aspect,
one often utilizes the concept of off-diagonal long-range
order (ODLRO) proposed by Yang as early as thirty years
ago\cite{s18}
In \cite{s18}, Yang showed that the existence of ODLRO of fermionic
systems imply Bose-Einstein condensation. This relationship was
made more clearer recently\cite{s19}\cite{s20}. It is indeed
supported
by  the BCS trial state, which does not  nevertherless belong
to the Hilbert space of the original Hamiltonian. Using the
symmetry
of the Hubbard model and $\eta$-pairing, Yang constructed
many eigenstates of the Hamiltonian
supporting ODLRO\cite{s2}. Later, Essler {\it et.al.}
showed that the ground states of a couple of generalized
Hubbard models possess ODLRO\cite{s4}\cite{s6}.
Though it is generally considered that the Hubbard model
may account for the high-$T_c$ enigma, the Kondo lattice model
and periodic Anderson model are also possibly relevent because
the superconducting properties are doping-dependent according
to experiments.\\
\indent In this letter, we  make use of the fluctuation-
dissipation theorem to study the pseudo-spins of the three models at
finite temperture. The prototype one-band
Hubbard model on an lattice $\Lambda$
\beq
H_{\rm HM}=\sum_{(ij)}\sum_{\sigma}t_{ij}c^{\dag}_{i\sigma}c_{j\sigma}+
U\sum_in_{i\uparrow}n_{i\downarrow}
\eeq
where $c^{\dag}_{i\sigma}$ and $c_{i\sigma}$ are the creation
and annihlation operators of the electrons with spin $\sigma=
\uparrow, \downarrow$ at site $i$. The hopping matrix
$\{t_{ij}\}$ are required
to be real and symmetric. The number operators are $n_{i\sigma}
=c^{\dag}_{i\sigma}c_{i\sigma}$, while the $U$ denotes the on-site
Coulomb interaction. It is further assumed that the the lattice
$\Lambda$
is bipartite in the sense that it
can be devided into sublattices {\bf A} and {\bf B}, i.e.
$\Lambda={\bf A}\cup{\bf B}$, such that $t_{ij}=o$ whenever
$\{ij\}\in {\bf A}$ or $\{ij\}\in {\bf B}$.
The Kondo lattice model is
\beq
H_{\rm KLM}=\sum_{(ij)}\sum_{\sigma}t_{ij}c^{\dag}_{i\sigma}c_{j\sigma}+
U\sum_in_{i\uparrow}n_{i\downarrow}
+J\sum_i{\bf S}^{loc}_i\cdot {\bf S}^c_i
\eeq
where ${\bf S}^{loc}$ are the localized spins of the impurities and
${\bf S}^c$ are the spins of the conduction electrons ( whose definition
will be given later). This model can be clearly regarded
as a doped Hubbard model. The periodic Anderson model
is described by
\beq
H_{\rm PAM}=\sum_{(ij)}\sum_{\sigma}t_{ij}c^{\dag}_{i\sigma}c_{j\sigma}+
U\sum_in_{i\uparrow}n_{i\downarrow}
+\sum_{i\sigma}\epsilon_f n^f_{i\sigma}
+V\sum_{i\sigma}(c^{\dag}_{i\sigma}f_{i\sigma}
+f^{\dag}_{i\sigma}c_{i\sigma})+U_f\sum_i n^f_{i\uparrow}
n^f_{i\downarrow}
\eeq
where $n^f_{i\sigma}=f^{\dag}_{i\sigma}f_{i\sigma}$ are the number
opertors of the localized electrons. Note that we have assumed
that the conduction electrons have also on-site Coloumb interaction
in both the Kondo lattice model and the periodic Anderson model.
When $U=0$, the model $H_{\rm PAM}$ is called {\it symmetric}
if $2\epsilon_f+U_f=0$. Here $U\not=0$, we call the model
symmetric if $2\epsilon_f+U_f=U$ and we consider this case only.\\
\indent For HM and KLM
the spin ${\bf S}_c$ and
pseudo-spin ${\bf \tilde{S}}_c$ for the conduction electrons
, which is equivalent to the
$\eta$-pairing, are defined as follows
\beq
S_c^{+}=\sum_{i\in\Lambda}c^{\dag}_{i\uparrow}c_{i\downarrow},
S_c^{-}=\sum_{i\in\Lambda}c^{\dag}_{i\downarrow}c_{i\uparrow},
S_c^z=\frac{1}{2}\sum_{i\in\Lambda}(c^{\dag}_{i\uparrow}
c_{i\uparrow}-c^{\dag}_{i\downarrow}c_{i\downarrow})=\frac{1}{2}
({\bf N}_{\uparrow}-N_{\downarrow})
\eeq
\beq
 \tilde{S}_c^+=\sum_{i\in\Lambda}\epsilon(i)C_{i\uparrow}
c_{i\downarrow}
,\,\,\,\,  \tilde{S}_c^-=\sum_{i\in\Lambda}\epsilon(i)
c^{\dag}_{i\downarrow}
c^{\dag}_{i\uparrow},\,\,\,\,\,
 \tilde{S}_c^z=\frac{1}{2}\sum_{i\in\Lambda}(1-n_{i\uparrow}
-n_{i\downarrow})
\eeq
where $\epsilon(i)=1 $when $i\in{\bf A}$ and $-1$ when
$i\in{\bf B}$
.Both the spin and the pseudo-spin operators
constitute SU(2) algebra and they
commute with each other, i.e. $[{\bf \tilde{S_c}}, {\bf S}_c]=0$ ,
so they form an SU(2)$\otimes$ SU(2) algebra.
For HM, the total spin is ${\bf S}=\sum_i {\bf S}^c_i$ while
for KLM, the total spin is ${\bf S}=\sum_i ({\bf S}^c_i+{\bf S}
^{loc}_i)$.  It is not difficult to show that  $[H, {\bf \tilde{S}}^2]=
[H, {\bf S}^2]=[H,\tilde{S}_z]=[H,  S_z]=0$.
Even, we have $[H,{\bf S}]=0$ (but $[H,{\bf \tilde{S}}]\not=0$
and $[H_{\rm KLM}, {\bf S}_c]\not=0$ ), $H=H_{\rm HM}, H_{\rm KLM}$.
.  So both HM and KLM enjoy SU(2)$\otimes$U(1)$\otimes$U(1)
symmetry. The spin is relevent to the
magnetic properties while the pseudo-spin is relevent to
superconducting. For HM, Yang and Zhang\cite{s3} showed that
ODLRO exists whenever the expectation value of
${\bf \tilde{S}}^2-
\tilde{S}_z^2$ is of order $N_{\Lambda}^2$, where
$N_{\Lambda}$
is the number of the sites of the lattice considered.\\
\indent For PAM, the spin operators and pseudo-spin
operators are\cite{s13}
\beq
S^+=\sum_i (c^{\dag}_{i\uparrow}c_{i\downarrow}+
f^{\dag}_{i\uparrow}f_{i\downarrow})
,\,\,\,\,\,\,\,\,
S^-=\sum_i (c^{\dag}_{i\downarrow}c_{i\uparrow}+
f^{\dag}_{i\downarrow}f_{i\uparrow})
\eeq
\beq
S^z=\frac{1}{2}\sum_i(c^{\dag}_{i\uparrow}
c_{i\uparrow}-c^{\dag}_{i\downarrow}c_{i\downarrow}
+f^{\dag}_{i\uparrow}f_{i\uparrow}
-f^{\dag}_{i\downarrow}f_{i\downarrow})
\eeq
\beq
\tilde{S}^+=\sum_i(-1)^i(c_{i\uparrow}c_{i\downarrow}
-f_{i\uparrow}f_{i\downarrow})
,\,\,\,\,\,\,\,
\tilde{S}^-=\sum_i(-1)^i(c^{\dag}_{i\downarrow}
c^{\dag}_{i\uparrow}-f^{\dag}_{i\downarrow}
f^{\dag}_{i\uparrow})
\eeq
\beq
\tilde{S}^z=\frac{1}{2}\sum_i(2-n_i-n^f_i)
\eeq
They both constitute SU(2) algebra.
\indent   The main result of this letter can be states as\\
{\it {\bf Theorem } For bipartite lattice $\Lambda$,
we have for HM, KLM. and symmetric PAM that
\beq
<{\bf \tilde{S}}^2-{\bf \tilde{S}}_z^2>(\mu, \beta)=<{\bf \tilde{S}}_z>
(\mu, \beta){\rm cth}
\beta\hbar(\frac{U}{2}-\mu)
\eeq
where $<O>(\mu, \beta)={\rm Tr}O\exp(-\beta K)/{\rm Tr}\exp(-\beta K),
K=H-\mu N, N=\sum_{i\sigma}n_{i\sigma}$ for HM and KLM
or $\sum_{i\sigma}(n_{i\sigma}+n^f_{i\sigma})$ for PAM.}\\
{\it Proof}. Consider the double-time Green function
$\ll{\bf \tilde{S}}^-
\mid{\bf \tilde{S}}^+\gg_{\omega}$. The evolution equation
of it is\cite{s14}
\beq
\omega\ll{\bf \tilde{S}}^-\mid{\bf \tilde{S}}^+\gg_{\omega}
=<[{\bf \tilde{S}}^-, {\bf \tilde{S}}^+]>+\ll[{\bf \tilde{S}}^-,
K]\mid{\bf \tilde{S}}^+\gg_{\omega}
\eeq
It can be calculated directly that
\beq
[{\bf \tilde{S}}^-,K]=(2\mu-U){\bf \tilde{S}}^-
\eeq
Accordingly, we have
\beq
\ll{\bf \tilde{S}}^-\mid{\bf \tilde{S}}^+\gg_{\omega}
=-\frac{2}{\omega+U-2\mu}<{\bf \tilde{S}}^z>
\eeq
Therefore, the expectation value $<{\bf \tilde{S}}^+
{\bf \tilde{S}}^->$
can be obtained by virtue of the fluctuation-dissipation
theorem
\beq
<{\bf \tilde{S}}^+{\bf \tilde{S}}^->=
\frac{i}{2\pi}\int^{+\infty}_{-\infty}\frac{
\ll{\bf \tilde{S}}^-\mid{\bf \tilde{S}}^+\gg_{\omega+i\eta}
-\ll{\bf \tilde{S}}^-\mid{\bf \tilde{S}}^+\gg_{\omega-i\eta}}
{\exp\beta\hbar\omega-1} d\omega
\eeq
Using $\lim_{\eta\rightarrow 0}\frac{1}{x\pm i\eta}={\cal P}
\frac{1}{x}\mp i\pi\delta(x)$, we have
\beq
<{\bf \tilde{S}}^+{\bf \tilde{S}}^->=-2<{\bf \tilde{S}}^z>
\frac{1}{\exp\beta\hbar(2\mu-U)-1}
\eeq
On the other hand, we have ${\bf \tilde{S}}^2
-{\bf \tilde{S}}^{2}_z
=\frac{1}{2}({\bf \tilde{S}}^+{\bf \tilde{S}}^-+
{\bf \tilde{S}}^-{\bf \tilde{S}}^+)$ and $[{\bf \tilde{S}}^+,
{\bf \tilde{S}}^-]=2{\bf \tilde{S}}_z$. Hence
\beq
<{\bf \tilde{S}}^2-{\bf \tilde{S}}^{2}_z>=
<{\bf \tilde{S}}^+{\bf \tilde{S}}^->-<{\bf \tilde{S}}_z>
\eeq
Using eq.(9), one can readily reach the conclusion.
\,\,\,\,\,\,\,  Q.E.D.\\
\indent Eq(15) can be employed to draw some conclusion on the electron
desities in the three models. Since $\tilde{S}^z=\frac{1}{2}(\kappa
N_{\Lambda}-N)$ where for HM and KLM, $\kappa=1$ while for PAM,
$\kappa=2$ ,we have
\beq
\frac{1}{N_{\Lambda}}<\tilde{S}^+\tilde{S}^->=(\kappa-\rho_e)\frac{1}
{1-e^{\beta\hbar(2\mu-U)}}
\eeq
where $\rho_e:=N/N_{\Lambda}$. From the finiteness of the l.h.s. of eq(17)
we have the following generalization of the lemma in \cite{s22}
\\
{\bf Corollary 1} {\it Under the same assumptions of the theorem, for all the
three models, we have:$\rho_e>\kappa$ if $2\mu>U$; $\rho_e=\kappa$ if
$2\mu=U$ and $\rho_e<\kappa$ if $2\mu<U$.}\\
\indent For PAM, we can write $\tilde{\bf S}=
\tilde{\bf S}_c+\tilde{\bf S}_f$.
Since for a finite PAM system, we alway have
\beq
<\tilde{\bf S}_c\cdot\tilde{\bf S}_f>=<\tilde{S}^z_c\tilde{S}^z_f
>=0
\eeq
because the bra and ket do not have equal number of creation
and annihilation operators of the same kindred. So we have the following\\
{\bf Corolary 2} {\it  For PAM,
\beq
<\tilde{\bf S}_c^2-\tilde{S}_c^{z2}>=
<{\bf \tilde{S}}_z>
{\rm cth}
\beta\hbar(\frac{U}{2}-\mu)-
<\tilde{\bf S}_f^2-\tilde{S}_f^{z2}> \geq 0
\eeq
}
\indent Since $\mu$ is the chemical potential (for PAM the
conduction electrons and the localized electrons must share
a common chemical potential because it is the total
electron number instead each of the respect species is
conserved), i.e.
the fermi energy in
the free fermion case, which is an intensive quantity,
it must be
a function of the particle density $\rho_e
=\frac{N_e}{N_{\Lambda}}$ and the
temperature $\beta$ and the model parameters,$t,U,...,
\mu=\mu(\rho_e,\beta,t,U,...,)$ where
$N_e$ is the number of electrons accommodated in the lattice.
Therefore, at any finite temperature,$\beta$, which is not a
root of $\frac{U}{2}-\mu(\rho_e,\beta,t,U,...,)=0$
the expectation value $<{\bf \tilde{S}}^2-
{\bf \tilde{S}}_z^2>$
can not be of order $N_{\Lambda}^2$ as $N_{\Lambda}
\rightarrow
\infty $ while keeping
$\rho_e$ constant according to theorem 1.
What about at the roots of $\frac{U}{2}-\mu(\rho_e,\beta,t,U,...,)=0$?
For the HM on a bipartite lattice, using the
complete particle-hole transform\cite{s23}
\beq
Pc_{i\uparrow}P^{-1}=\epsilon(i)c^{\dag}_{i\uparrow},
\,\,\,\,\,\,\,
Pc_{i\downarrow}P^{-1}=\epsilon(i)c^{\dag}_{i\downarrow}
\eeq
For KLM, the complete particle-hole transform is defined by (19) together
with
\beq
PS_i^{\pm loc}P^{-1}=-S_i^{\mp loc},\,\,\,\,\,\,\,
PS_z^{loc}P^{-1}=-S_z^{loc}
\eeq
i.e., the particle-hole transform effects a rotation of the spins
. As to for PAM, the transform is defined by (19) together with
\beq
Pf_{i\sigma}P^{-1}=-\epsilon(i)f_{i\sigma}^{\dag},\,\,\,\,\,
Pf_{i\sigma}^{\dag}P^{-1}=-\epsilon(i)f_{i\sigma},\,\,\,\,\,
\eeq
One has for all the three models
\beq
PKP^{-1}=H+\kappa(U-2\mu)N_{\Lambda}+(\mu-U)N
\eeq
Then since
\beq
<N>_{\mu,\beta}=\frac{1}{Z}\frac{1}{\beta}\frac{\p}{\p\mu}{\rm Tr}
e^{-\beta K}\\\\
=\frac{1}{Z}\frac{1}{\beta}\frac{\p}{\p\mu}{\rm Tr}
e^{-\beta[H+U(\kappa N_{\Lambda}-N)-\mu(2\kappa N_{\Lambda}-N)]}
=2\kappa N_{\Lambda}-<N>_{-(\mu-U), \beta}
\eeq
Therefore  we always have $<N>=\kappa N_{\Lambda}$
whenever $\mu=U/2$ no matter at what temperature.
So  $\mu=U/2$ can not determine the temperature.
(This is why we often say the system is at half-filling when $\mu=U/2$.
for HM). We have
accordingly $<\tilde{S}_z>=0$ in this case and the $r.h.s.$ of (10)
is in fact $0/0$, which should be determined by the limit $\lim_{\mu
\rightarrow U/2}$. Since
\beq
\frac{1}{\beta}\frac{\p}{\p\mu}<N>=<N^2>-<N>^2
\eeq
we have
\beq
\lim_{\mu\rightarrow U/2}
<{\bf \tilde{S}}^2-{\bf \tilde{S}}_z^2>=\frac{1}{2}(<N^2>-<N>^2)
\eeq
Since for an ideal gas, the $r.h.s.$ of (25), which
is just the fluctuation squared of particle number,
is of order $O(N)$, it is quite reasonable to assume the
in the three models here, it is also of the order $O(N)$. Hence
we have therefore a stronger conclusion than that in\cite{s24}.\\
{\it {\bf Corollary 3}. Under the same assumptions of the
theorem, the $l.h.s.$ of (10) can be at most of order
$O(N_{\Lambda})$ at any finite temperature for HM, KLM and PAM.\\}
\indent As in ref.\cite{s25}, eq(23) can be used to obtain a symmetry of the
l.h.s. of eq(10). Since eq(24) states that
\beq
\rho_e(\mu, \beta)=2\kappa-\rho_e(U-\mu, \beta)
\eeq
we have immediately from (17) that
\beq
\frac{1}{N_{\Lambda}}<\tilde{S}^+\tilde{S}^->(\mu,\beta)
=\frac{1}{N_{\Lambda}}<\tilde{S}^+\tilde{S}^->(U-\mu,\beta)\exp
\{\beta\hbar(U-2\mu)\}
\eeq
There exists lot of forms of the l.h.s. of (17) satisfying this symmetry.
For instance, $C(\beta, U)e^{-\beta\hbar(\mu-U/2)}$ and $C(\beta, U)\frac{1}
{1+e^{-\beta\hbar(U-2\mu)}}$ (where $C(\beta, \mu)$ denotes come function)
. So the functional form can not be
determined uniquely without further constraints.\\
\indent By virtue of particle-hole transform
(for PAM, $\epsilon$ is to be repalce by $(-1)^i$)
\beq
Tc_{i\uparrow}T^{-1}=\epsilon(i)c^{\dag}_{i\uparrow},\,\,\,\,\,
Tc_{i\downarrow}T^{-1}=c_{i\downarrow}
\eeq
\beq
Tf_{i\uparrow}T^{-1}=-\epsilon(i)f^{\dag}_{i\uparrow},\,\,\,\,\,
Tf_{i\downarrow}T^{-1}=f_{i\downarrow}
\eeq
under which the spin and pseudo-spin operators transform
into each other\cite{s9}.
\beq
T(S^+,  S^-,  S_z)T^{-1}
=(\tilde{S}^+,  \tilde{S}^-,  \tilde{S}_z)
,\,\,\,\,\,\,\,
T(\tilde{S}^+, \tilde{S}^-, \tilde{S}_z)T^{-1}
=(S^+,  S^-,  S_z)
\eeq
One can obtain the knowledge of spin of the transformed
model from the known knowledge of pseudo-spin of a given model\cite{s24}.
From the theorem we know that
at half filling, $<\tilde{S}_z>=0$, therefore
we always have $<{\bf \tilde{S}}^2-\tilde{S}_z^2>=0$
at any $\beta\not=\beta_c$, this agrees with the
Corolary 2 in \cite{s5}. It was shown that\cite{s22} at half-filling
for HM, no ODLRO is exhibited for on-site electron pairs in the translational
invariant case for both positive and negative $U$.
Yet, away from half-filling,
it seems that theorem 1 disagrees
with the theorem 1 of \cite{s9} since that
theorem
asserts that for negative $U$ Hubbard model and some special $\rho_e$,
the expectation value $<{\bf \tilde{S}}^2-
\tilde{S}_z^2>$ at
ground state can be of order $N_{\Lambda}^2$.
But the special $\rho_e$ was given {\it ad hoc} and was not
determined by the grand canonical equilibrium.\\
\indent As a by-product of the complete particle-hole transform,
we have also the following\\
{\it {\bf Corollary 4}. For the three models considered under the same
assumptions as in the theorem, at half-filling, i.e. $\mu=U/2$, one has
$<S_{x,z}>=0$ }\\
{\it Proof} Since under the complete particle-hole transform, $PS_{x,z}
P^{-1}=-S_{x,z}$, we have
\beq
<S_{x,z}>=\frac{1}{Z}{\rm Tr}S_{x,z} e^{-\beta K}
=\frac{1}{Z}{\rm Tr}(-S_{x,z})\exp\{-\beta[H+\kappa(U-2\mu)N_{\Lambda}
+(\mu-U)N]\}
\eeq
Therefore at $\mu=U/2$, we have
\beq
<S_{x,z}>=-<S_{x,z}>
\eeq
Hence one can obtain the assertion.
\vskip 0.3in
\underline{Acknowledgement} The author expresses sincere
gratitude to Prof. Y.Q. Li for helpful discussions. Also is the author
indebted to Prof. S. Randjbar-Daemi for his invitation for three
months at ICTP. This
work
was supported by the Funds for Young Teachers of Shanghai Education
Committee and in part by the National Science Foundation under
Grant. 19805004.\\

\end{document}